\documentclass[aps,twocolumn,groupedaddress,showpacs]{revtex4}
\usepackage{graphicx}

\begin{document}

\title[Mesoscopic ring]
{The large system asymptotics of persistent currents in mesoscopic
quantum rings}

\author{A.~Gendiar$^{1,2}$, R.~Krcmar$^{1}$, and M. Weyrauch$^3$}
\affiliation{$^1$ Institute of Electrical Engineering, Slovak Academy of Sciences,
SK-841~04, Bratislava, Slovakia\\
$^2$ Institute for Theoretical Physics C, RWTH University Aachen, D-52056 Aachen, Germany\\
$^3$ Physikalisch-Technische Bundesanstalt, D-38116 Braunschweig, Germany}

\date{\today}

\begin{abstract}
We consider a one-dimensional mesoscopic quantum ring filled with
spinless electrons and threaded by a magnetic flux, which carries a
persistent current at zero temperature. The interplay of Coulomb
interactions and a single on-site impurity yields a non-trivial
dependence of the persistent current on the size of the ring. We
determine numerically the asymptotic power law for systems up to
$32\,000$ sites for various impurity strengths and compare with
predictions from Bethe Ansatz solutions combined with Bosonization.
The numerical results are obtained using an improved functional
renormalization group (fRG) method. We apply the density matrix
renormalization group (DMRG) and exact diagonalization methods
to benchmark the fRG calculations. We use DMRG to study the
persistent current at low electron concentrations in order to
extend the validity of our results to quasi-continuous systems.
We briefly comment on the quality of calculated fRG ground state
energies by comparison with exact DMRG data.
\end{abstract}


\maketitle

\section{Introduction}

Quantization of the magnetic flux was first observed in
superconducting cylinders~\cite{Deaver, Byers}. In normal electronic
systems the flux quantum $\phi_0=hc/e$  determines the period of the
persistent currents, which can be observed in mesoscopic
one-dimensional disordered metallic rings~\cite{Buttiker}.
Persistent currents in metallic rings are induced only if the
circumference of the ring is not larger than the electron phase
coherence length and/or the electron's mean-free path. Such
conditions can be easily obtained for sufficiently low temperatures
at which inelastic scattering from phonons is suppressed. Normally,
this happens at temperatures below 1~K and ring sizes of about
1~$\mu$m. Impurities (disorder) in such metallic rings act as
elastic scatterers, and it is not surprising that persistent
currents can flow despite the non-zero resistance of the rings.

Interest in persistent current phenomena remains strong both
experimentally~\cite{Levy, Chandrasekhar, Mailly, Reulet, Jariwala,
Rabaut, Kleemans, Bouchiat} and theoretically
~\cite{Landauer, Cheung, Abraham, Bouzerar, Gogolin, Kato, Cohen,
Henkel, Jaimungal, Meden, Enss, Friedrich, Soroker, Heinrichs}. The
main reason for this is the fact that there still is a discrepancy
between experiment and theory. Experiments ~\cite{Levy, Jariwala}
show persistent currents which are two orders of magnitude larger
than theoretical predictions for rings filled with non-interacting
electrons~\cite{Cheung, Altshuler, Henkel}. If electron-electron
interactions are included, persistent currents increase for
repulsive as well as attractive electron interactions~\cite{Abraham,
Ambegaokar, Gogolin}, however, the calculated currents were still
about five times smaller than the experimental data. Recently,
Bary-Soroker {\it et al.}~\cite{Soroker} provided an explanation
for the magnitude of the persistent currents if magnetic impurities
and the attractive interactions are considered.

Considerable theoretical effort has been invested into the question
whether the interplay between strong electron-electron interactions,
electron density, and disorder strength can explain the large
persistent currents observed experimentally. Disorder usually gives
rise to elastic scattering and consequently should reduce the
persistent current. Bouzerar {\it et al.}~\cite{Bouzerar} claim that
both electron-electron interactions and disorder  decrease the
persistent current, whereas the self-consistent Hartree-Fock
calculations by Cohen {\it et al.}~\cite{Cohen} suggest
that the current may be enhanced by a moderate disorder due to
screening effects. Other theoretical studies conclude that
electron-electron interaction could reduce or enhance the current
depending on the type of disorder present in the
system~\cite{Abraham, Cesare, Gambetti}. Abraham and
Berkovits~\cite{Abraham}  conjectured that the increase would be
rather small and one-dimensional spinless models at half-filling
cannot explain the results of single ring experiments.

Among the theoretical methods used to study persistent current
phenomena are Hartree-Fock approximations~\cite{Kato,Cohen},
configuration interaction and quantum Monte Carlo
simulations~\cite{Krcmar,Nemeth}, Bosonization~\cite{Gogolin},
conformal field theory~\cite{Henkel,Jaimungal}, the functional
renormalization group (fRG)~\cite{Meden}, and the density matrix
renormalization group (DMRG)~\cite{Meden,Cesare,Gambetti}.

In the present paper we study the interplay between
electron-electron interactions and a single impurity in a
one-dimensional system. For one-dimensional interacting systems,
the presence of a single impurity is known to affect their
physical properties~\cite{Luther,Apel,Kane1,Kane2}. We restrict
our investigations to the Luttinger liquid phase which is
characterized by a power law decay of correlation functions
in a sufficiently large system~\cite{Haldane}. Our aim is to
investigate the asymptotic behavior of the persistent current
$I$ as a function of the ring length $L$, from which the power
law exponent may be obtained $I\propto L^{-\alpha_{\rm B}-1}$.
The exponent $\alpha_{\rm B}=K^{-1}-1$ is related to the Luttinger
liquid parameter $K$~\cite{Haldane, Kane1, Kane2} and depends on
the electron-electron interaction, but it does not depend on the
impurity strength if $L\to\infty$.

It is known that the asymptotics of the currents is reached for
smaller system sizes if sufficiently strong impurities are
considered~\cite{Gogolin,Matveev,Glazmann}. Nevertheless, it is
necessary to investigate relatively large systems in order to reach
the asymptotics. We therefore need suitable many-body techniques
which enable such calculations. We have chosen the functional
renormalization group~\cite{Polchinski,
Wetterich,Morris,Salmhofer1,Salmhofer2}, which proved to be a rather
useful tool in a previous study~\cite{Meden}. However, for numerical
reasons the asymptotic regime could not really be reached in that
work. In the present paper we improve the fRG technology such that
system sizes up to about 32000 lattice sites can be investigated.
In order to benchmark the fRG results we use DMRG
calculations~\cite{White1,White2,Peschel,Scholl}.

We also study the asymptotic power law decay of the persistent
current at low electron concentrations in order to clarify
differences between discrete lattice models (investigated in this
paper) and continuous systems studied by quantum Monte Carlo and
configuration interaction methods~\cite{Nemeth, Krcmar}. For this
reason, we apply DMRG to our model away from half-filling
($n=\frac{1}{2}$), in particular, we study the case of
quarter-filling ($n=\frac{1}{4}$) as well as $n=\frac{1}{8}$
and $n=\frac{1}{16}$.

The paper is organized as follows. In Sec. 2 we define the model for
our calculations and briefly review the theory of the persistent
current. We introduce the fRG method in Sec.~3 and related
Appendices. Section~4 is devoted to the DMRG method with a short
discussion  of its limitations. Results are presented in Sec.~5
gathering the numerical data obtained by fRG, DMRG, and exact
diagonalization. We calculate the persistent current and extract the
effective exponent $\alpha_{\rm eff}$ related to the power law
decrease with respect to the ring size $L$. Finally we show DMRG
results for systems with low-filling. The results are summarized in
Sec.~6.

\section{Model for spinless Fermions}

We consider a quantum ring of interacting spinless electrons at
zero temperature. The ring is pierced by a magnetic flux $\phi$
and contains one single impurity. This system is modelled by the
tight-binding Hamiltonian
\begin{equation}\label{hamiltonian}
{\cal H}=-t\sum\limits_{\ell=1}^{L}
\left(
c^\dagger_{\ell}c^{}_{\ell+1}e^{-i\phi/L}+{\rm h.c.}
\right)
+U\sum\limits_{\ell=1}^{L}n_{\ell}n_{\ell+1}+V\,n_{1}\, ,
\label{eq1}
\end{equation}
written in terms of the electron creation and annihilation operators
$c^\dagger_\ell$ and $c_\ell$ as well as the electron density
operator $n_\ell=c_\ell^\dagger c_\ell$.

The Peierls factor $e^{-i\phi/L}$ with $\phi=2\pi\Phi/\Phi_0$
describes the influence of the magnetic field in terms of the
magnetic flux $\Phi$~\cite{Peierls}. The flux quantum $\Phi_0=h c/e$
is set to one in the following. The hopping amplitude $t$ between
neighboring sites is also set to one in order to define the energy
scale. The nearest-neighbor Coulomb interaction is denoted by $U$.
An (external) on-site impurity potential with strength $V$ is
placed on the  lattice site $\ell=1$ along the ring
($\ell=1,2,...,L$). Periodic boundary conditions are imposed by
$L+1\equiv1$.

\begin{figure}[tb]
\centerline{\includegraphics[width=0.47\textwidth,clip]{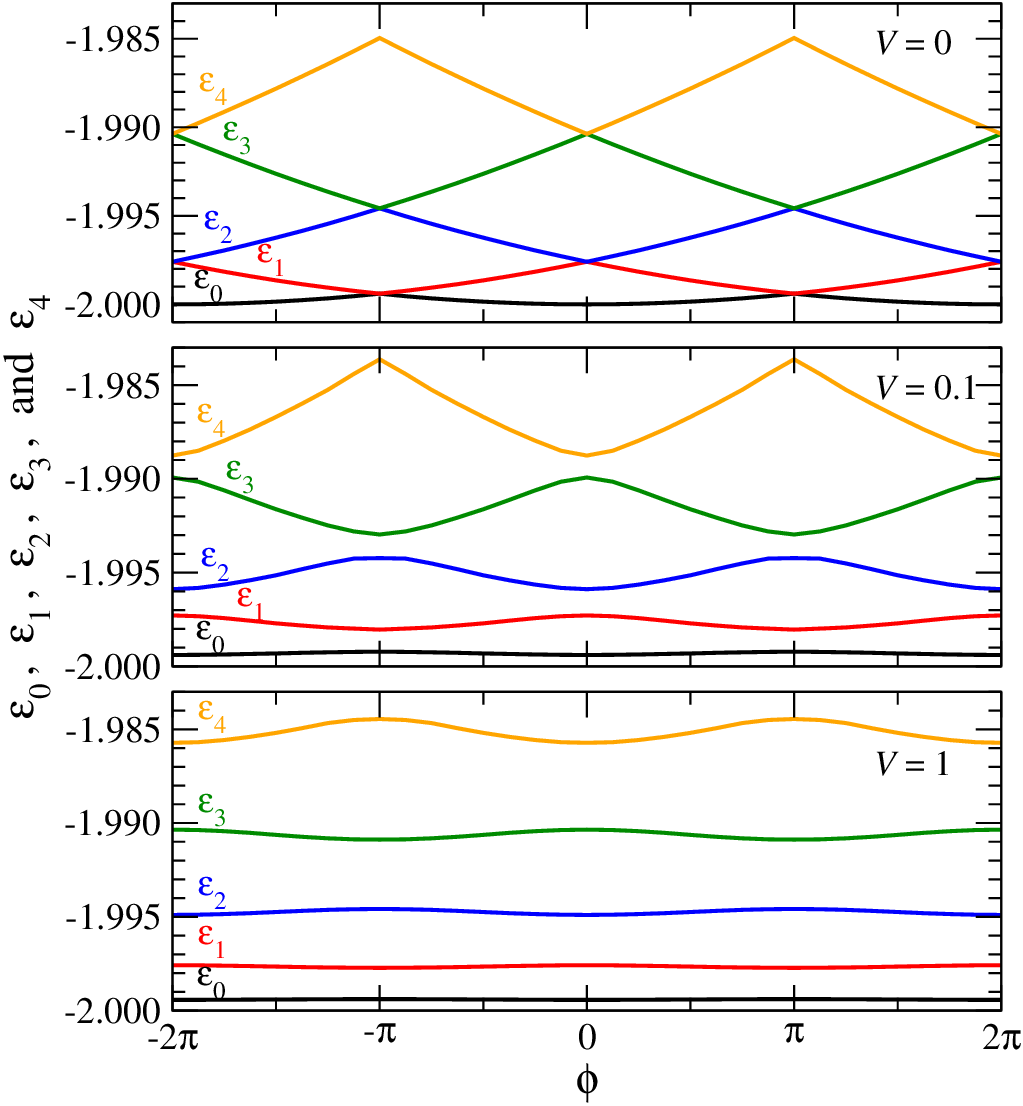}}
\caption{(Color online) The five lowest-laying single-particle energy
levels for a non-interacting system as functions of the magnetic flux
$\phi$ for a ring of size $L = 128$. The three panels correspond to
$V=0$, $V=0.1$, and $V=1.0$, respectively.}
\label{f1}
\end{figure}

Without electron-electron interaction ($U=0$) and impurity ($V=0$)
the single particle energy levels $\epsilon_m$ calculated from the
Hamiltonian Eq.~(\ref{hamiltonian}) depend quadratically on the
magnetic flux $\phi$. The persistent current $I_m=-v_m / L$ at
energy level $m$ with the wave vector $k=2\pi \left(\phi+m\right)/L$
is calculated from the electron velocity $v_m$,
\begin{equation}
v_m(\phi)=2\pi\frac{\partial \epsilon_m(\phi)}{\partial k(\phi)}
=L\frac{\partial \epsilon_m(\phi)}{\partial\phi}\, .
\end{equation}
Summation over all occupied energy levels then yields the total
persistent current at temperature $T=0$,
\begin{equation}\label{percurrent}
I(\phi)=-\sum_{m=0}^{m_{\rm max}}\frac{\partial \epsilon_m(\phi)}{\partial\phi}\, .
\end{equation}

The upper panel in Fig.~\ref{f1} shows the parabolic energy band
structure ($V=0,~U=0$). Consequently, the persistent current is
proportional to the magnetic flux $\phi$ and  inversely proportional
to the length $L$ of the ring. For an odd number of electrons in the
ring one finds
\begin{equation}
I(\phi)=-\frac{v_F}{\pi L}\phi\, ,\qquad-\pi\leq\phi<\pi\, ,
\label{nc1}
\end{equation}
whereas for an even filling the current is given by
\begin{eqnarray}
I(\phi)=-\frac{v_F}{\pi L}\times\left\{
\begin{array}{ll}
\phi+\pi\, ,\qquad & -\pi\leq\phi<0,\\
\\
\phi-\pi\, ,\qquad & 0\leq\phi<\pi
\end{array}
\right.
\label{nc2}
\end{eqnarray}
with $v_{\rm F}$ being the Fermi velocity.

An impurity in the ring rounds off the cusps seen in this function
as shown in the central and lower panels of Fig.~\ref{f1} for a weak
and an intermediate impurity, respectively. For strong impurities
and odd filling, one finds $I\propto-{v_{\rm F}}\sin(\phi)/{\pi
L}$~\cite{Gogolin}. The coupling strength $V$ of an on-site
impurity can be related to the physically more relevant transmission
coefficient at the Fermi wave vector $k_{\rm F}$
\begin{equation}
T(k_{\rm F})=\frac{4t^2}{4t^2+V^2}\, ,
\end{equation}
which holds at half-filling in the non-interacting limit~\cite{SCH98}.
With this relation, the strengths of various impurities $V$ used in
this paper may be compared.

\begin{figure}[tb]
\centerline{\includegraphics[width=0.47\textwidth,clip]{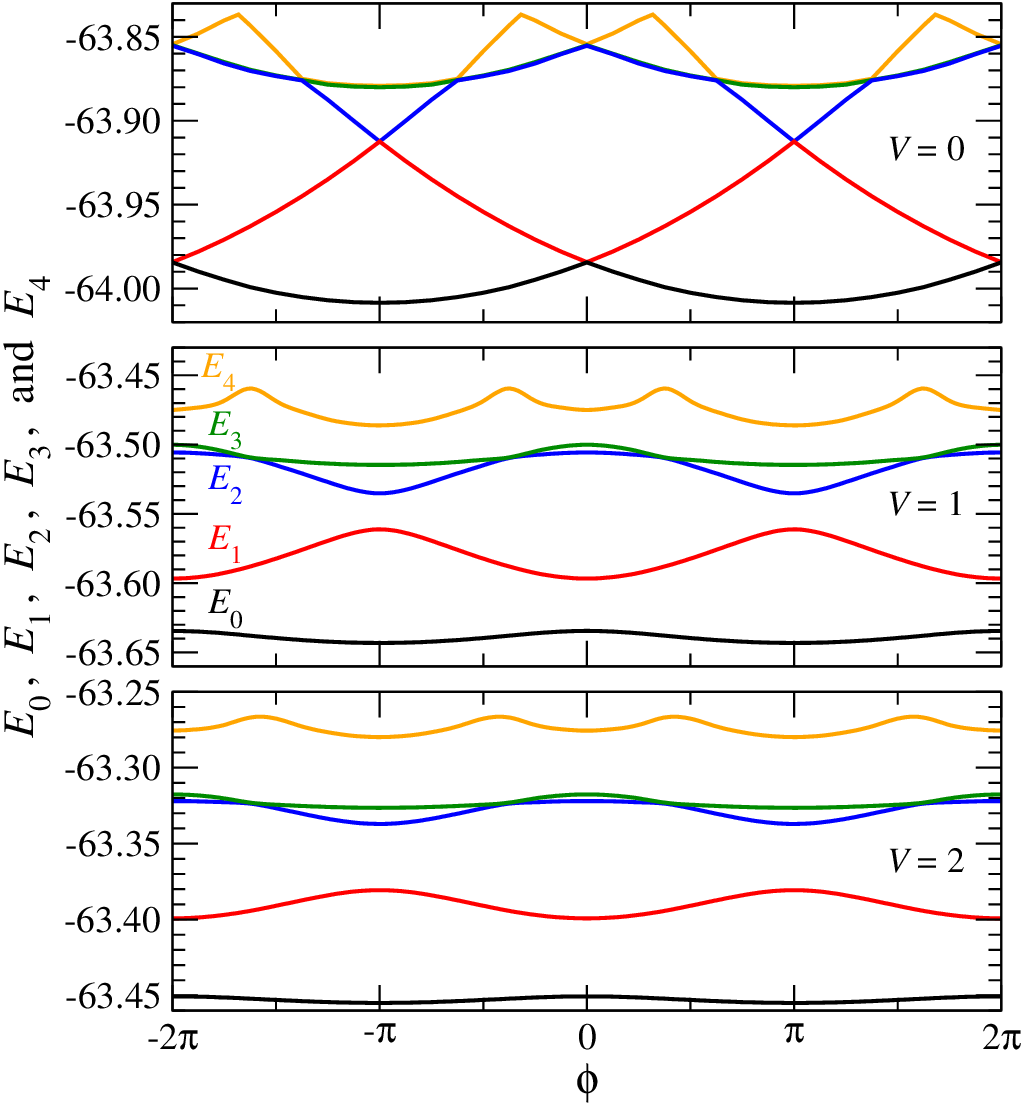}}
\caption{(Color online) The five lowest-laying many-body energy levels
obtained by DMRG as a function of the magnetic flux $\phi$ for a ring
of size $L = 128$ at half-filling and electron-electron interaction
$U=1$. Upper panel: $V=0$, center panel: $V=1$, lower panel: $V=2$.}
\label{f2}
\end{figure}

If the electron-electron interaction $U$ is switched on, the
many-body energy levels are shown in Fig.~\ref{f2} for $U=1$
and various impurity strengths $V$. The persistent currents
are obtained from the relation
\begin{equation}\label{pcf}
I(\phi)=-\frac{\partial E_0(\phi)}{\partial\phi}\, ,
\end{equation}
where $E_0(\phi)$ is the many-body ground-state energy at zero
temperature as plotted in Fig.~\ref{f2}. In general the system
is a Luttinger liquid. However, at half-filling it is in the
Luttinger liquid phase only for $|U|\leq 2$. At $U=2$, the system
exhibits a charge-density wave instability, and phase separation
characterizes the system for $U<-2$.

In this paper we focus on the Luttinger liquid phase for which
Bosonization techniques and additional approximations predict
that asymptotically (i.e. for large $L$) the persistent current
decays algebraically with increasing system size,
\begin{equation}\label{alpha}
I\propto L^{-1-\alpha_{\rm B}}\, .
\end{equation}
The exponent $\alpha_{\rm B}$ is a function of the electron-electron
interaction (regardless of the impurity strength $V$). Only
at half-filling, the exponent $\alpha_{\rm B}$ can be obtained
analytically from a Bethe Ansatz solution~\cite{Haldane,Meden,And}
\begin{equation}
\alpha_{\rm B}=\frac{2}{\pi}\arccos\left(-\frac{U}{2}\right)-1\, .
\end{equation}
Throughout this paper we often consider the case $U=1$ with the
corresponding $\alpha_{\rm B}= \frac{1}{3}$. We determine
$\alpha_{\rm B}$ from the Hamiltonian~Eq. (\ref{hamiltonian})
using three different many-body techniques: fRG, DMRG, and
exact diagonalization (ED).

\section{Functional RG scheme for interacting Fermions}\label{FRG}

A scheme for one-dimensional Fermionic systems based on the
functional renormalization group, has been developed by Meden,
Metzner, Sch{\"o}nhammer, and collaborators~\cite{MED02a} based on
seminal work by Wetterich~\cite{Wetterich} and Morris~\cite{Morris}.
It has been applied to study transport properties of quantum
wires~\cite{MED02a,MED02b,AND04} and rings~\cite{Meden}.

The fRG scheme of the present work is similar to the one applied in
the references cited above, but uses a somewhat different truncation
procedure. Details of our method have been presented
recently~\cite{WEY08}. Here we review the essential steps relevant
for a system of spinless interacting Fermions.

The effective average action $\Gamma_k$~\cite{Wetterich} for
interacting Fermions evolves according to the fRG
equation \cite{Wetterich, BER00}
\begin{equation}\label{method-fRG}
\frac{\partial}{\partial k}\Gamma_k [\phi^*,\phi]=-\frac{1}{2}{\rm
Tr}
    \left\{ [{\Gamma}_k^{(2)}[\phi^*,\phi]+R_k]^{-1}\frac{\partial {R}_k}{\partial
    k}\right\}.
\end{equation}
The system of $L$ spinless electrons is described by an
$L$-component vector of Grassmann variables
$\phi(\tau)=(\phi_1(\tau),\ldots,\phi_L(\tau))$, which describes the
evolution of the interacting Fermionic system in imaginary time
$\tau$.  The functional derivative of $\Gamma_k$ with respect to
$\phi^*$ and $\phi$ is denoted by $\Gamma_k^{(2)}$. The trace in
Eq.~(\ref{method-fRG}) is to be performed over the Fermionic states
$j=1,\ldots,L$. The regulator ${R}_k$ is introduced in order to
suppress thermal and quantum fluctuations at energy or momentum
scales $k$ larger than any physical scale relevant for our problem.
With decreasing $k$, the regulator gradually ``switches on" such
fluctuations until they are fully included at $k=0$, i.e. at $k=0$
the regulator $R_k$ vanishes. As a regulator we choose
$R_k=Ck\theta(k^2-\omega^2)$ with $C$ a large constant and it
satisfies all requirements for a useful regulator as discussed in
more detail in Ref.~\cite{WEY08}. It has an additional advantage
that the integration over the frequencies $\omega$ can be done
analytically.

In order to solve the functional differential
equation~(\ref{method-fRG}) a particular functional form for
$\Gamma_k$ in terms of the Grassmann variables $\phi^*, \phi$ is
assumed
\begin{equation}\label{method-Gamma}
\Gamma_k[\phi^*,\phi]=\int_0^\beta {\rm d}\tau
\sum_{\alpha=1}^L\phi_\alpha^*(\tau)\frac{\partial}{\partial\tau}\phi_\alpha(\tau)+
       \mathcal{U}_k\left(\phi^*(\tau),\phi(\tau)\right),
\end{equation}
where the `effective potential' $\mathcal{U}_k$ does not depend on
derivatives of the Grassmann variables with respect to the imaginary
time $\tau$, i.e. it is represented as a Grassmann polynomial in
terms of $\phi^*$ and $\phi$
\begin{eqnarray}
 \mathcal{U}_k(\phi^*,\phi)&=&a_{k,0}+\sum_{j=1}^L a_{k,jj}\phi_j^*\phi_j
+a_{k,j,j+1}\phi_j^*\phi_{j+1}\nonumber\\
&+&a_{k,j,j-1}\phi_{j}^*\phi_{j-1}\nonumber\\
& +&U_k\sum_{j=1}^L\phi^*_j\phi_j\phi^*_{j+1}\phi_{j+1}
\end{eqnarray}
with the understanding that $\phi_0=\phi_L$ and $\phi_{L+1}=\phi_1$.
Of course, in general this polynomial should have higher order
terms, but those are neglected in the hope they are small.

For half-filling, the density-density interaction strength $U_k$
renormalizes according to
\begin{equation}
U_k=\frac{U}{1+\frac{U}{2\pi}\left(k-\frac{2+k^2}{\sqrt{4+k^2}}\right)}
\label{ren_U}
\end{equation}
as was suggested in Ref.~\cite{And}. This result can be easily
derived within the fRG scheme applied here by using analogous
approximations for the two-particle vertex as in Ref.~\cite{And}.
(The interaction parameter $U$ is given in the Hamiltonian
Eq.~(\ref{hamiltonian})).

Inserting this Ansatz for the effective average action on both sides
of the flow equation, performing the integration over the frequencies
$\omega$, and comparing terms with the same Grassmann structure,
it is straightforward to obtain the flow equations for the coefficients
$a_{k,0}$ and $a_{k,jj^\prime}$
\begin{eqnarray}\label{floweqs}
a_{k,0}^\prime&=&\frac{1}{2\pi}\sum_{\lambda=\pm k}e^{i\lambda0^+}
\ln {\rm det}\left(\frac{1}{i\lambda}\mathbf{g}_k^{-1}(i\lambda)
\right),\nonumber\\
a_{k,jj}^\prime&=&\frac{U_k}{2\pi}\sum_{\lambda=\pm k}\sum_{s=\pm 1}
e^{i\lambda0^+}g_{k,(j+s)(j+s)}(i\lambda),\nonumber\\
a_{k,j(j\pm 1)}^\prime&=&-\frac{U_k}{2\pi}\sum_{\lambda=\pm k}
e^{i\lambda0^+}g_{k,j(j\pm 1)}(i\lambda)
\end{eqnarray}
with $\mathbf{g}_k(i\lambda)=\left(\mathbf{a}_k+i\lambda
\mathbf{1}\right)^{-1}$ and $\mathbf{a}_k=(a_{k,jj^\prime})$.
The convergence factor $e^{i\lambda0^+}$ is only needed at the
beginning of the flow from infinity down to a large constant $k_0$.

At $k=\infty$ the effective average action $\Gamma_k$ is given by
the classical action ${\cal S}$~\cite{Wetterich, BER00}, which is
determined by the Hamiltonian Eq.~(\ref{hamiltonian}). Therefore,
the initial conditions at $k=\infty$ for the solution of the flow
equations~(\ref{floweqs}) are directly obtained from
Eq.~(\ref{hamiltonian}),
\begin{eqnarray}\label{initialcond}
a_{\infty,0}&=&0,\nonumber\\
a_{\infty,jj}&=&-V\delta_{j1}+\mu\delta_{jj}, \\
a_{\infty,j(j\pm 1)}&=&e^{\mp i\phi/L}, \nonumber
\end{eqnarray}
with the understanding that $a_{L,L+1}=a_{L,1}$ and
$a_{1,0}=a_{1,L}$. For convenience we have added a chemical
potential term $-\mu n_\ell$ to the Hamiltonian. Our initial
conditions differ from those of Ref.~\cite{Meden}: in the present
work the dependence on the magnetic flux $\phi$ resides in the
initial conditions for the off-diagonal $a_{jj^\prime}$ only, while
in Ref.~\cite{Meden} this dependence partly resides in the free
propagator, corresponding to a different definition of the kinetic
energy. Moreover, here we also include the chemical potential into
the $a_{jj^\prime}$ and not in the free propagator.

In practice we must start the renormalization flow at a finite
$k=k_0$. The initial flow from $k=\infty$ to $k=k_0$ is obtained
from an analytical solution of the Eqs.~(\ref{floweqs}). For
convenience this calculation is briefly reviewed in
Appendix~\ref{initial cond} with the result
\begin{eqnarray}\label{incondeff}
a_{k_0,0}&=&\frac{V}{2}+\frac{L}{2}\left(\frac{U}{2}-\mu\right), \nonumber\\
a_{k_0,jj}&=&(\mu-U)\delta_{jj}-V\delta_{j1},\\
a_{k_0,j(j\pm 1)}&=&e^{\mp i\phi/L}.\nonumber
\end{eqnarray}
At half-filling the chemical potential is given by $\mu=U$ because
of particle-hole symmetry. From the  ground state energy
$E_0(\phi)=a_0(k=0)+\mu\langle n \rangle$, the persistent current is
calculated using Eq.~(\ref{percurrent}).

For given $L$, Eq.~(\ref{floweqs}) represents a (large) set of
$3L-1$ non-linear coupled complex-valued ordinary differential
equations. The calculation of the right-hand side of these equations
requires the inversion of a (potentially) large cyclic tridiagonal
matrix as well as the computation of the logarithm of the
determinant of such a matrix at each step of the numerical
integration of the differential equations. To accomplish this in an
efficient manner is described in some detail in
Appendix~\ref{numerics}.

\section{DMRG}\label{DMRG}

The density matrix renormalization group is a numerical technique
for the diagonalization of the very large matrices typically
encountered in quantum many-body calculations. The technique is
described in detail in Refs~\cite{White1,White2, Peschel,Scholl}. We
use DMRG for the calculation of the ground-state energy $E_0(\phi)$.
In our case, the matrices to be diagonalized are complex-valued due
to the Peierls factor $e^{-i\phi/L}$ entering Eq.~(\ref{eq1}).
Treating such complex-valued systems by DMRG does not lead to
numerical instabilities, apart from a few rare cases which occur at
very low electron fillings. Here, however, the numerical
diagonalization routines for complex matrices experience convergence
difficulties for very large system sizes. Also, the superblock
diagonalization routines seldom require more than $10^4$ cycles for
convergence as compared to the standard $\sim10^2$ cycles typically
necessary for real-valued matrices. The memory requirements increase
by a factor of $1.8$ and the calculation time raises by a factor of
$2.1$ compared to a standard real-valued DMRG.

We kept the DMRG truncation error as small as $\varepsilon <
10^{-9}$ for system sizes $L\geq128$, for smaller systems
$\varepsilon \approx 10^{-15}$ could be achieved. The number of
states kept are left to vary such that the above truncation
error condition could be satisfied. It is well known that the
efficiency and accuracy of DMRG decreases substantially if periodic
boundary conditions are imposed. We checked the accuracy of our
results comparing with data obtained from the exact diagonalization
of the Hamiltonian for systems with up to $L=24$ sites at various
fillings.

We would like to point out that the calculations for strong impurities
$V>10$ and for large rings with $L>100$ require an extremely careful
treatment because differences between the ground state energies for
different magnetic flux values are extremely small $|E_0(\phi=0)
-E_0(\phi=\pi)|/|E_0(\phi=0)|<10^{-6}$. For this reason, both the
number of states kept and the number of the finite-size method
sweeps must be sufficiently large.

\section{Results}\label{results}

In this section we present and analyze numerical data obtained
by three different methods: fRG, DMRG, and ED. We
study the dependence of the persistent currents on the ring size
$L$, the impurity strength $V$, the magnetic flux $\phi$, and on the
electron concentration $n$. Varying the electron-electron
interaction $U$ within the Luttinger liquid regime $-2\leq U\leq2$
does not change physical properties qualitatively as we shall see
below. However, the weaker the electron-electron interaction the
larger is the system size $L$  needed to reach the asymptotic power
law decay of the persistent current~\cite{Meden, Krcmar}.

The ground state energies $E_0(\phi)$ are calculated for
$0\leq\phi\leq\pi$ and extended to the first Brillouin zone
$-\pi\leq\phi\leq\pi$ using the reflection symmetry of
the energy around the origin $\phi=0$ (cf. Figs.~\ref{f1} and
\ref{f2}). The persistent current in Eq.~(\ref{pcf}) obtained by
DMRG and ED is calculated using numerical differentiation.
The Fourier coefficients $I_k$ of the currents
\begin{equation}
I(\phi)=\sum_{k=1}^\infty I_k \sin(k\phi)
\label{FA}
\end{equation}
can be calculated without numerical differentiation using an
integration by parts
\begin{equation}
I_k=\frac{2k}{\pi}\int_0^\pi {\rm d}\phi E_0(\phi)\cos(k\phi).
\end{equation}
Nevertheless, small numerical errors in the calculation of
the ground state energy may significantly affect the calculation
of the power law exponent $\alpha_B$ (cf.~Eq.~(\ref{alpha})).
The Fourier analysis for strong impurity strengths in
Eq.~(\ref{FA}) is well justified due to the sine-like shape
of the current, and the higher coefficients $I_k$, $k\ge2$
decay to zero with increasing $L$ faster than $I_1$. However,
if the impurity $V\ll1$, the first Fourier coefficient $I_1$
may not suffice to characterize the decay of the persistent
current accurately.

\begin{figure}
\centerline{\includegraphics[width=0.47\textwidth,clip]{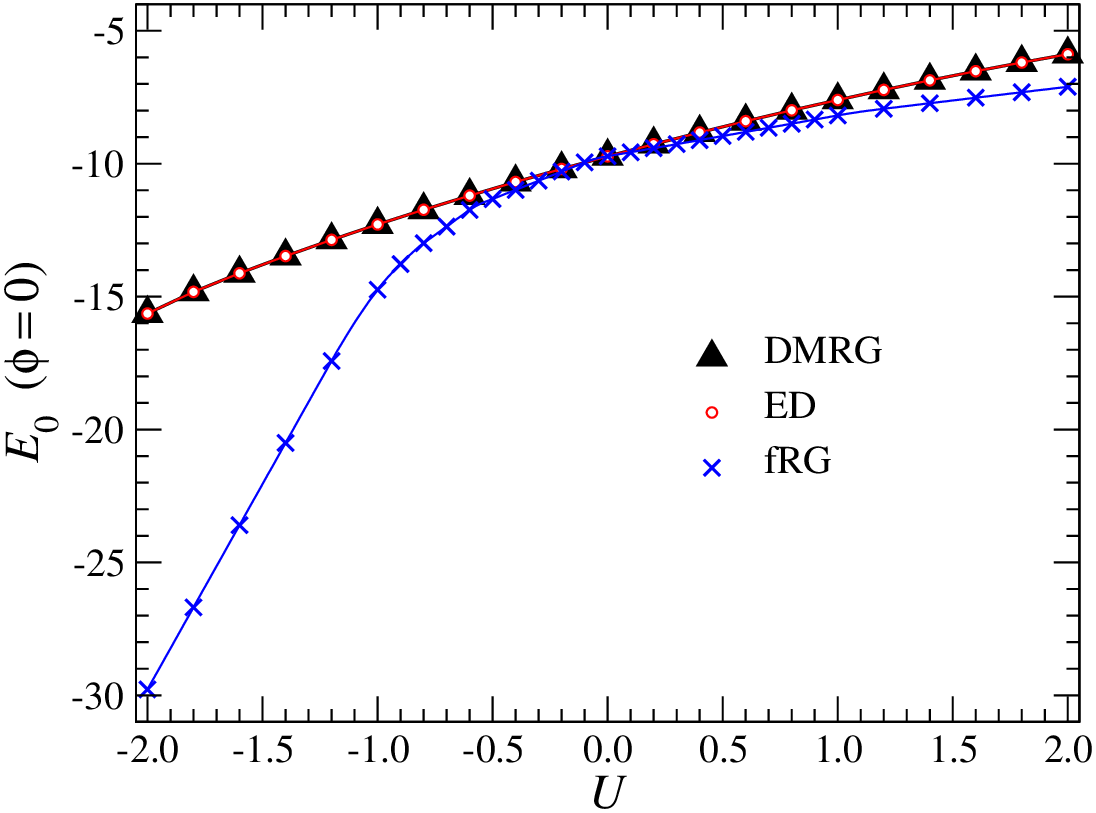}}
\caption{(Color online) Ground state energies at $\phi=0$
calculated by DMRG (filled triangles), ED (open circles), and
fRG (crosses) as functions of $U$ within the Luttinger liquid
regime for $L=16$ and $V=1$.}
\label{fig17}
\end{figure}

\begin{figure}
\centerline{\includegraphics[width=0.47\textwidth,clip]{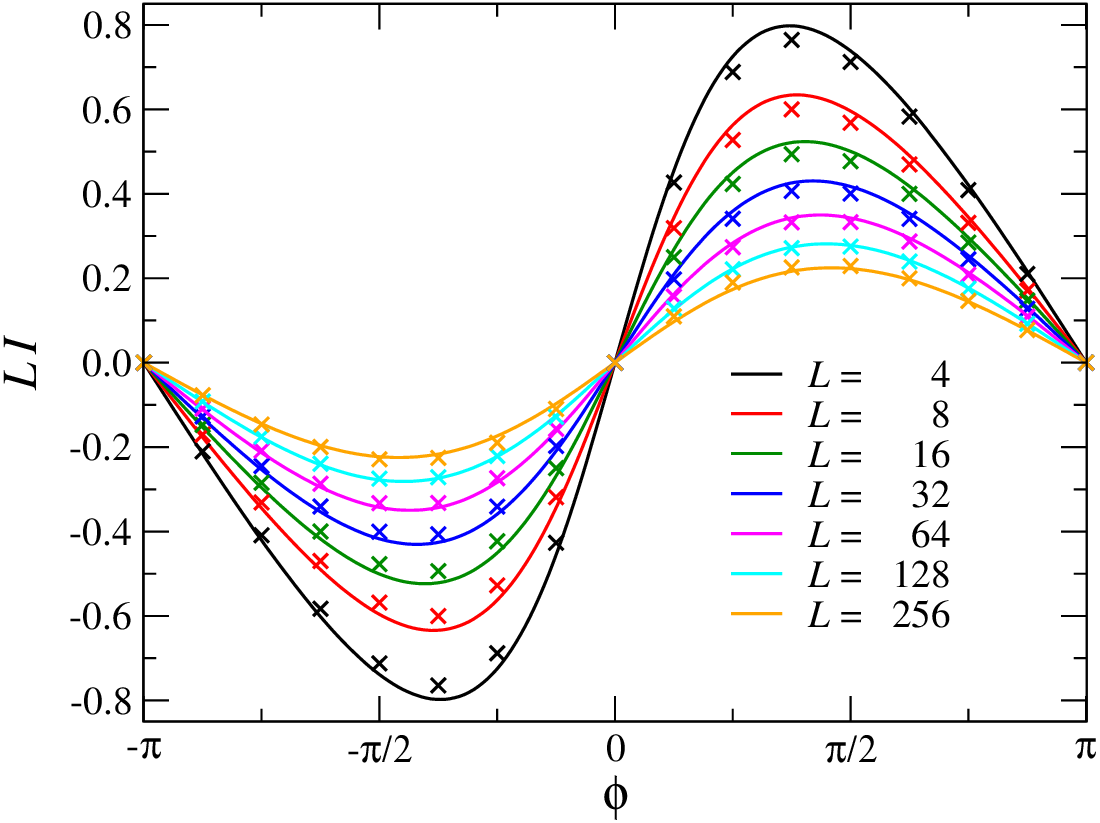}}
\caption{(Color online) Persistent currents calculated by fRG (lines)
and DMRG (symbols $\times$) at half-filling for  $U=1$, $V=2$, and various
ring lengths $L$. The currents with the largest amplitude correspond
to $L=4$ and then the amplitudes decrease with increasing $L$.}
\label{pcfdmrg}
\end{figure}

For system sizes $L\leq24$ and for impurities $V<100$, ED and DMRG
yield essentially exact ground state energies in mutual agreement.
The ground state energy calculated from fRG differs from
the exact results of ED and DMRG at large interactions, as
shown in Fig.~\ref{fig17} for interactions within the Luttinger
liquid regime. This deviation signals the drastic approximations
involved in the fRG method used here, in particular, the discarding
of higher order vertex functions. However, the shape of the ground
state energy as a function of the magnetic flux can be reproduced
quite well by this method as can be seen from Fig.~\ref{pcfdmrg}.
Therefore, we can use fRG calculations in order to obtain the
persistent currents for systems as large as $L=3\cdot 10^4$.
For very strong impurities ($V\gg10$) DMRG and fRG may run into
numerical difficulties and we analyzed such cases by ED as
is discussed in more detail below.

\begin{figure}
\centerline{\includegraphics[width=0.47\textwidth,clip]{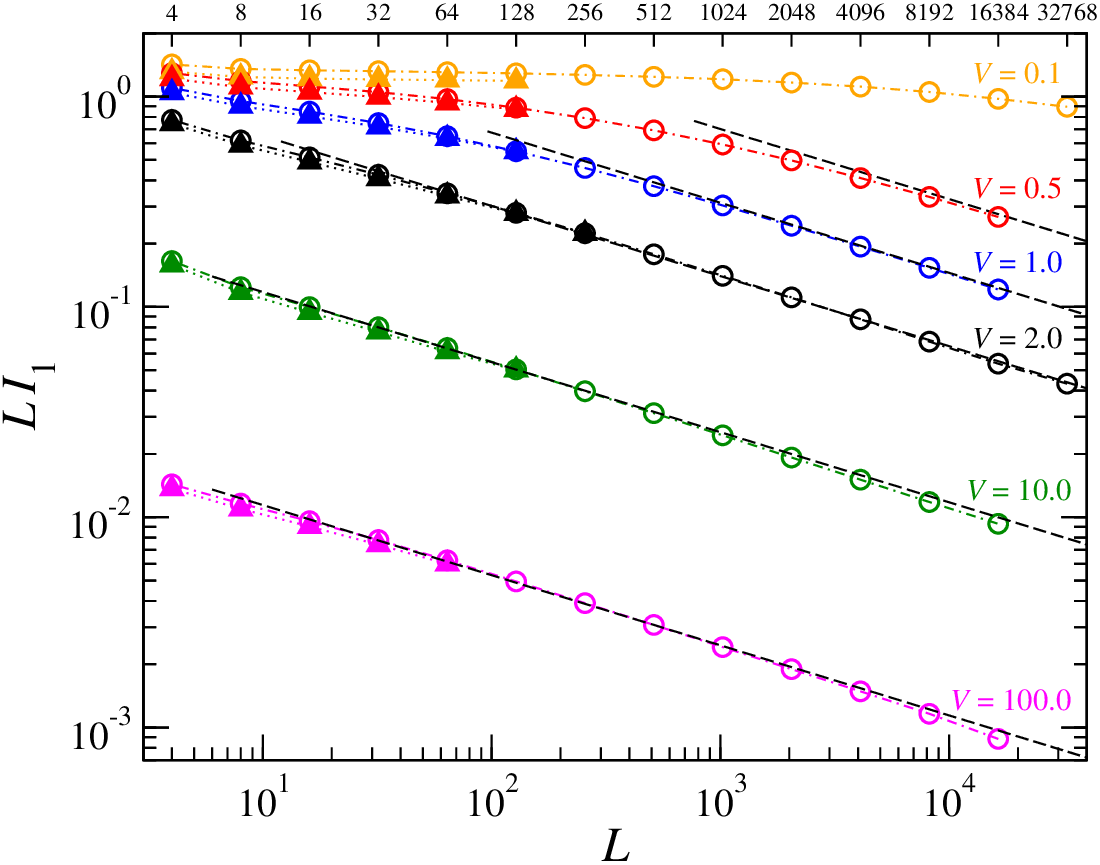}}
\caption{(Color online) The first Fourier coefficient
$I_1$ versus the length of the ring $L$ for $U=1$ and various $V$.
The open circles and the filled triangles, respectively, correspond
to the fRG and DMRG data. The dashed straight lines show the asymptotics
$\alpha_{\rm B}=\frac{1}{3}$. The upper $x$-axis shows the exact number
of the sites $L$ at which the calculations were performed.}
\label{ffc}
\end{figure}
\begin{figure}
\centerline{\includegraphics[width=0.47\textwidth,clip]{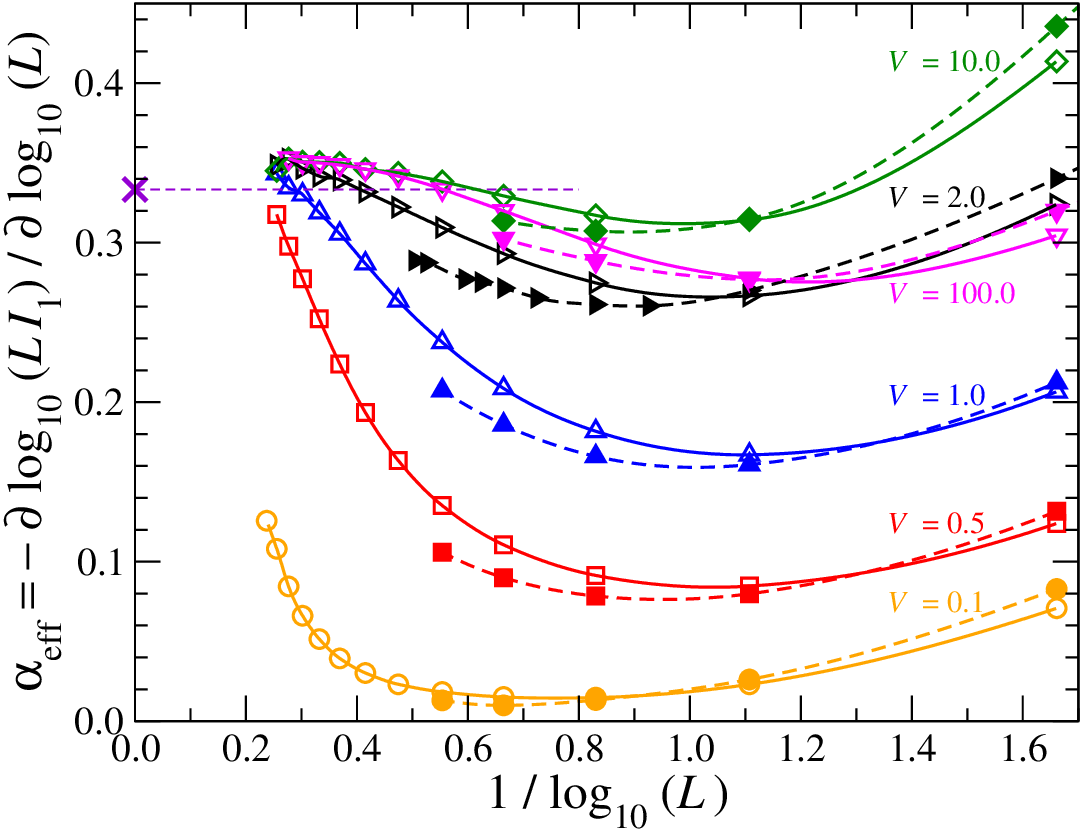}}
\caption{(Color online) The effective exponent $\alpha_{\rm eff}$
versus the inverse of $\log_{10}(L)$ for the interaction $U=1$ and
various impurities $V$. The open and filled symbols correspond to
fRG and DMRG data, respectively.}
\label{alphaeff}
\end{figure}

\subsection{fRG and DMRG analysis at half-filling}

Figure~\ref{pcfdmrg} shows persistent currents as functions of the
flux calculated at half-filling ($n=\frac{1}{2}$) using DMRG and fRG
for various ring lengths $L$ and for an intermediate impurity
strength $V=2$ (corresponding to transmission $T=\frac{1}{2}$).
The data indicate that for small and intermediate
ring lengths $L$, the DMRG and fRG calculations agree rather well.
DMRG calculations are hardly feasible for $L>256$, and only fRG data
are available in order to study the large $L$ limit.

Bosonization theory including additional approximations predict
a power law decay of the persistent current at large $L$,
cf.~Eq.~(\ref{alpha}). In Fig.~\ref{ffc} we compare this asymptotic
decay (dashed lines) with the numerically determined first Fourier
coefficient of the persistent current $LI_1$ in a logarithmic plot.
From Bosonization one expects that $LI_1$ decays asymptotically
as $L^{-\alpha_{\rm B}}$. The open circles represent the fRG
data and the filled triangles the DMRG results. Again we see the
good agreement between both methods, and it is obvious that the
asymptotic decay does not depend on the impurity strength (if at least
$V>0.1$), which ranges from rather weak $V=0.1$ ($T=0.9975$) to very
strong $V=100$ ($T=0.0004$). The asymptotic regime is reached at
smaller $L$ for stronger impurities.

Notice a tiny deviation of the fRG data from the dashed lines
with the common power-law exponent $\alpha_{\rm B}$ at large
$L$. This is due to the approximations involved in the fRG method
and is known from studies of quantum wires~\cite{Enss}. However,
there are also numerical limitations: for large systems the
differences between currents calculated for different magnetic
fluxes are so tiny that they can not be resolved by the floating
point data representation of the computer, and the current cannot be
calculated reliably. This numerical limitation also prevents us to
calculate even larger systems.

In order to quantify the deviation of the numerical data from
the expected power law, we define an effective exponent
\begin{equation}
\alpha_{\rm eff}=-\frac{\partial\log_{10}(LI_1)}
{\partial\log_{10}(L)}\, ,
\label{ae}
\end{equation}
which is shown in Fig.~\ref{alphaeff} as a function of
$1/\log_{10}(L)$. The data labeled by the open and filled symbols are
obtained from the fRG and DMRG results shown in Fig.~\ref{ffc}
using a numerical derivative. If the impurity is weak ($V=0.1$), the
effective exponent $\alpha_{\rm eff}$ deviates significantly from
$\alpha_{\rm B} =\frac{1}{3}$ (the horizontal dashed line). It is
expected that substantially larger systems would need to be
considered to approach the asymptotics in this case. For $V\geq0.5$,
the numerically determined exponents tend to saturate at
$\alpha_{\rm eff}\approx0.35$ which is about 5\% larger than the
expected $\alpha_{\rm B}=\frac{1}{3}$. Our result is in
good agreement with the leading $U$-behavior, where the correction
to the exponent for the quantum wire is known to be
$\frac{U}{\pi}=0.318$~\cite{Enss}. As already stated, we
attribute this deviation to the approximations involved in the
fRG. Similar results were obtained in a study on open
chains~\cite{AND04}, where the same power law decay is extracted
from the decay of the spectral weight near the impurity site.

From our data analysis it emerges that the effective exponents for
larger $L$ are also non-negligibly influenced by the numerical
limitations discussed above, which are seen in the figure as small
irregularities in the approach of the numerical data to saturation.
Notice that we are dealing with extremely tiny effects observed in
the exponent $\alpha_{\rm eff}$. Thus, extremely accurate
calculations of the ground state energy is the key to the
analysis of the effective exponent. There is another feature of the
effective exponent worth mentioning: it shows a characteristic
minimum at lengths $8 \lesssim L \lesssim 100$ within the range
of $V$.

The effective exponent determined from the DMRG data (full symbols)
is systematically below the fRG results for all but the very small
systems. Since the DMRG data are expected to be accurate, one may
conjecture that $\alpha_{\rm eff}$ approaches $\alpha_{\rm
B}=\frac{1}{3}$ for $L\to\infty$, if DMRG calculations for such
large systems would be feasible.

\begin{figure}[tb]
\centerline{\includegraphics[width=0.47\textwidth,clip]{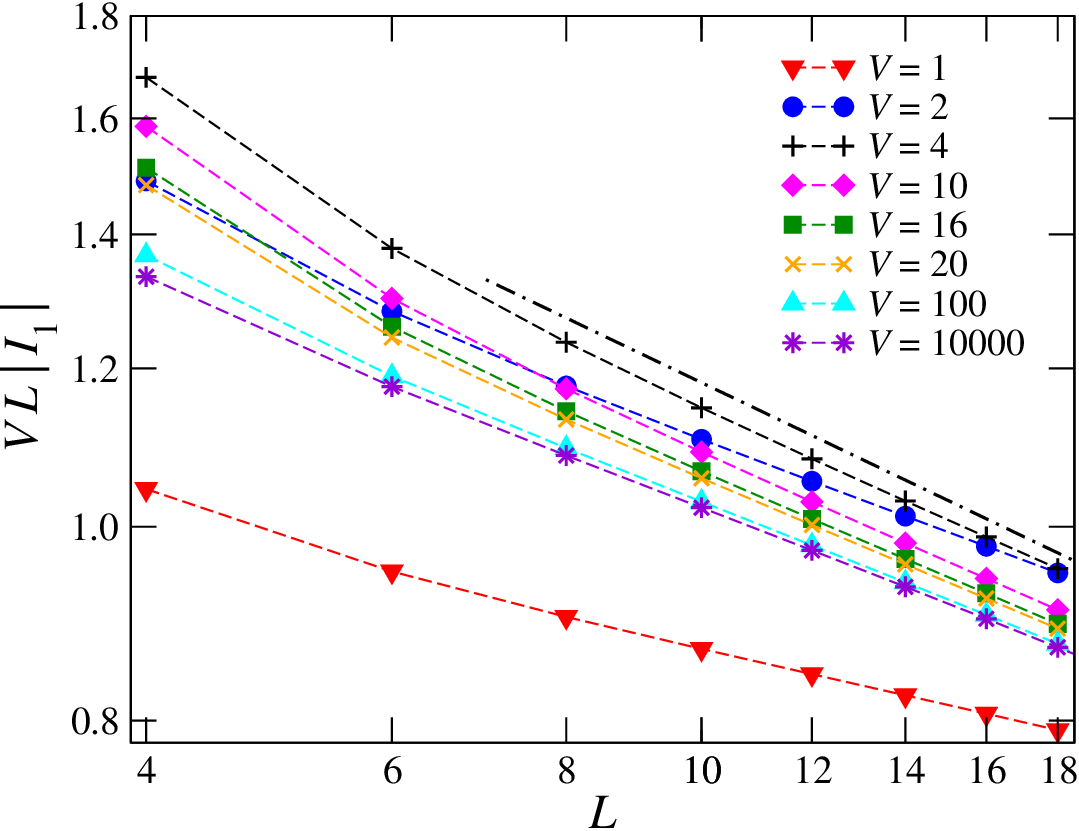}}
\caption{(Color online) The persistent current versus $L$ for a variety
of $V$'s obtained by ED at $U=1$ and half-filling.}
\label{EDpc}
\end{figure}
\begin{figure}
\centerline{\includegraphics[width=0.47\textwidth,clip]{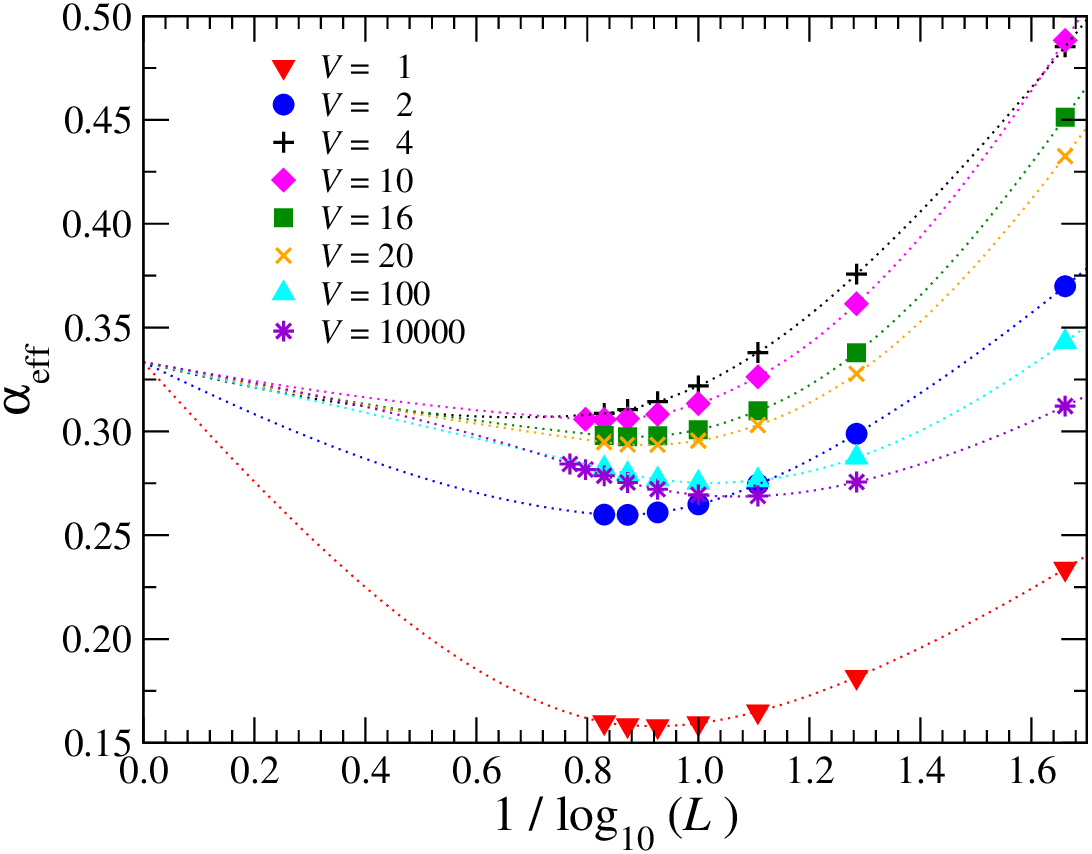}}
\caption{(Color online) Dependence of the effective exponent on
$1/\log_{10}(L)$ for the same parameters as in Fig.~\ref{EDpc}.}
\label{EDea}
\end{figure}
\begin{figure}
\centerline{\includegraphics[width=0.47\textwidth,clip]{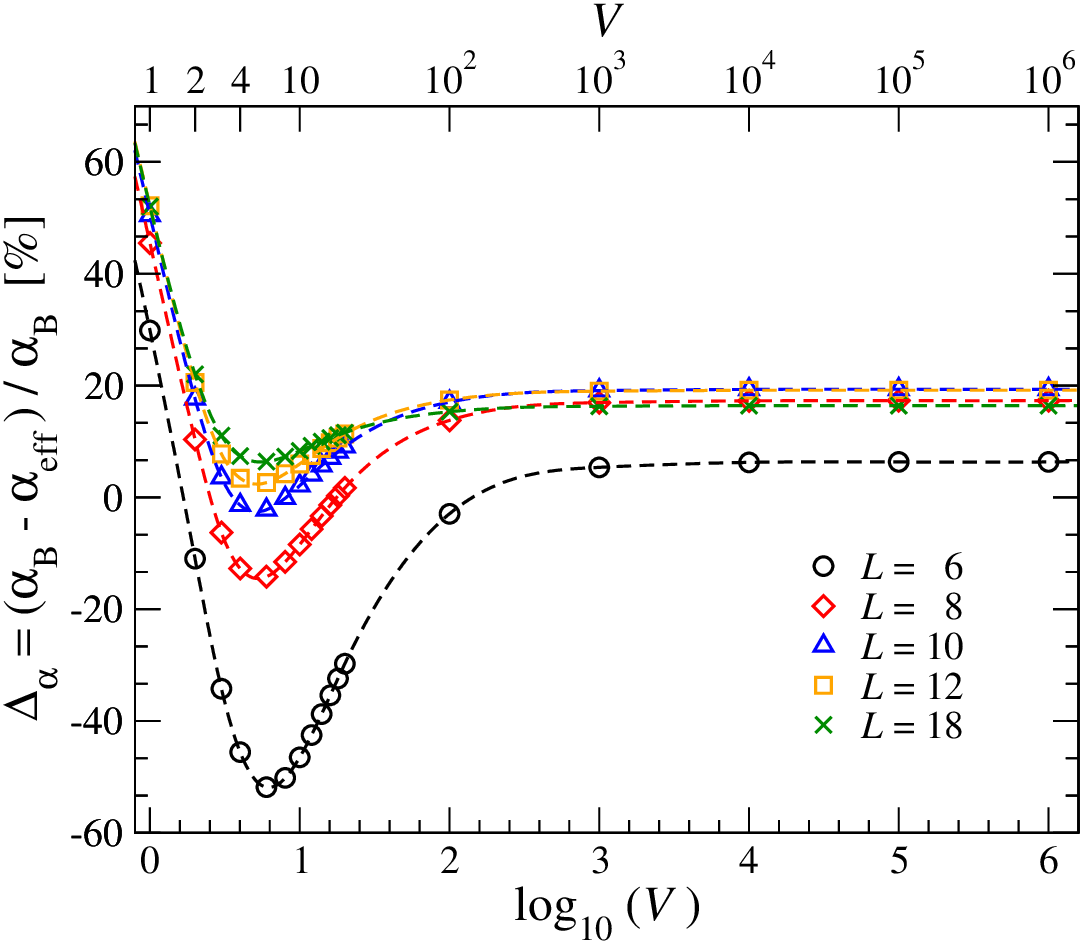}}
\caption{(Color online) Relative difference between the effective
and asymptotic exponents as a function of the impurity strength
$V$ for several sizes $L$.}
\label{EDrd}
\end{figure}

\subsection{Non-monotonous behavior of the effective exponent}

Another interesting feature can be seen in Fig.~\ref{alphaeff}. Both
fRG and DMRG data show that the result for the strongest impurity
$V=100$ appears to be out of sequence if compared to the other
$V$'s. To elucidate this further, we performed a series of
calculations using ED at smaller sizes $L$. Figure~\ref{EDpc}
shows the first Fourier coefficient of the persistent current
$I_1$ as a function of $L$ on a logarithmic scale for on-site impurities
in the range $1\leq V\leq 10^4$. In order to gather the data,
the Fourier coefficient is scaled with $VL$ in Fig.~\ref{EDpc}.
Furthermore, we plot the absolute value of the Fourier coefficient
in order to remove the even-odd effect discussed in Section~II.
The straight dot-dashed line corresponds to $\alpha_{\rm B}=\frac{1}{3}$.

Figure~\ref{EDea} shows corresponding calculations of the effective
exponent $\alpha_{\rm eff}$ for the same set of impurity strengths.
The dotted lines only serve as guides to the eye and should not be
taken as extrapolations. Again, it is obvious that the data for the
cases $V \ge 10^2$ appear to be out of sequence as in Fig.~\ref{alphaeff}.
In Fig.~\ref{EDrd} we plot the relative difference $\Delta_{\alpha}
=(\alpha_{\rm B}-\alpha_{\rm eff})/\alpha_{\rm B}$ in order to
quantify the dependence of the effective exponent on the impurity
strength $V$ for several system sizes $L$. A minimum of
$\Delta_{\alpha}$ appears at $V\approx4$ for $U=1$. We,
therefore, identify three regions in the figure (for reachable
system sizes): in the first region, $0<V\lesssim4$, we observe
the standard behavior, i.e., the stronger the impurity, the faster
the convergence of $\alpha_{\rm eff}$ to $\alpha_{\rm B}$. In the
the second region which starts at around $V\approx4$ and ends
around $V=10^3$, this behavior is reversed. In the third region
$V>10^3$ the effective exponent $\alpha_{\rm eff}$ saturates and
does not change any more with increasing impurity strength.

\subsection{DMRG study at low fillings}\label{low filling}

\begin{figure}
\centerline{\includegraphics[width=0.47\textwidth,clip]{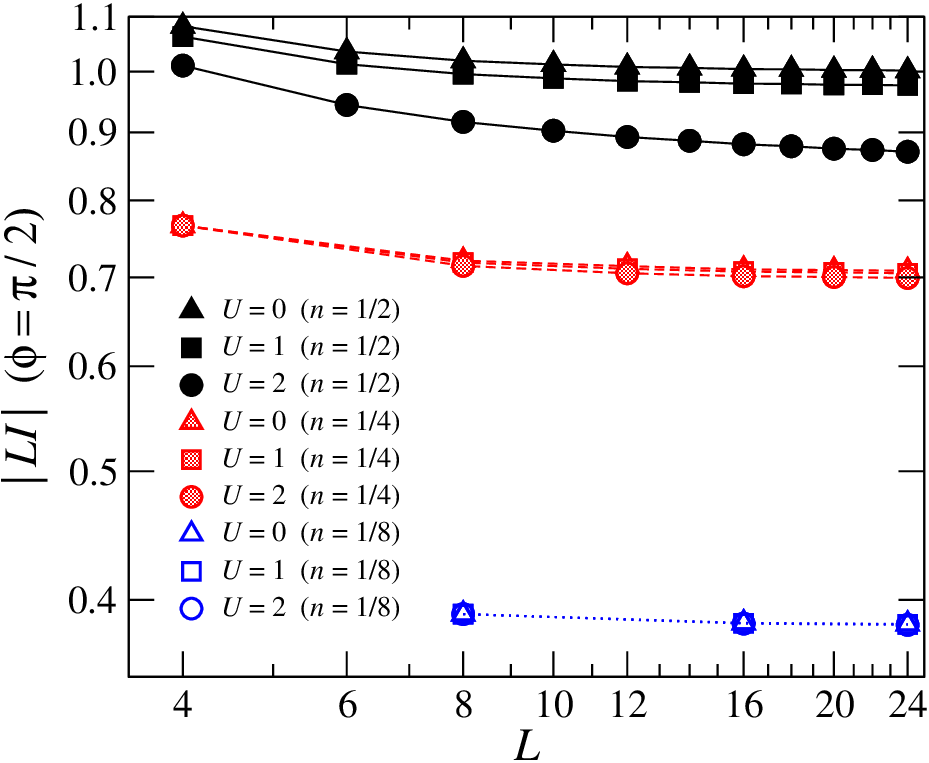}}
\caption{(Color online) Persistent currents as a function of the
system size $L$ for magnetic flux $\phi=\frac{\pi}{2}$,
various electron concentrations $n$, and interactions $U$.
Plotted is the absolute value of the persistent current in order to remove
the sign change between even and odd electron fillings.}
\label{galinv}
\end{figure}
\begin{figure}
\centerline{\includegraphics[width=0.47\textwidth,clip]{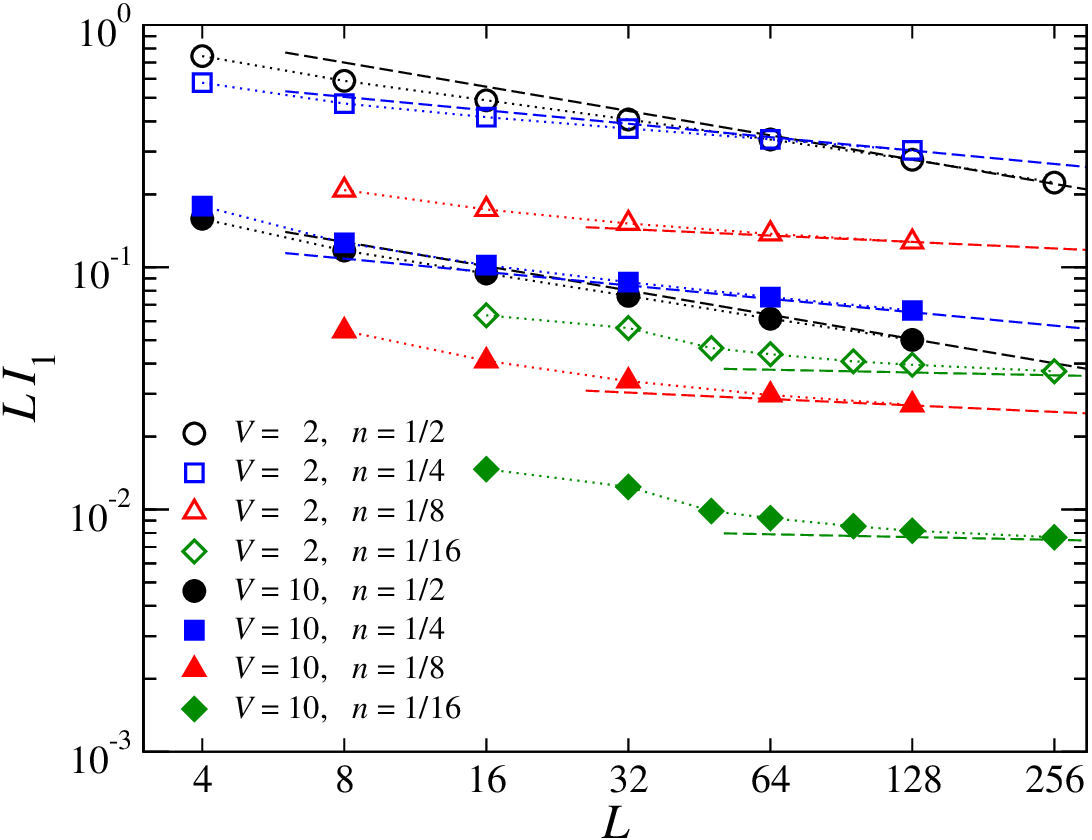}}
\caption{(Color online) Persistent currents versus system size  $L$
for $U=1$ at various electron fillings $n$. The open and full
symbols correspond to the impurity $V=2$ and
$V=10$, respectively. The slopes of the dashed lines coincide with
the asymptotic exponents $\alpha_{\rm B}$.}
\label{pcgi}
\end{figure}

\begin{figure}
\centerline{\includegraphics[width=0.47\textwidth,clip]{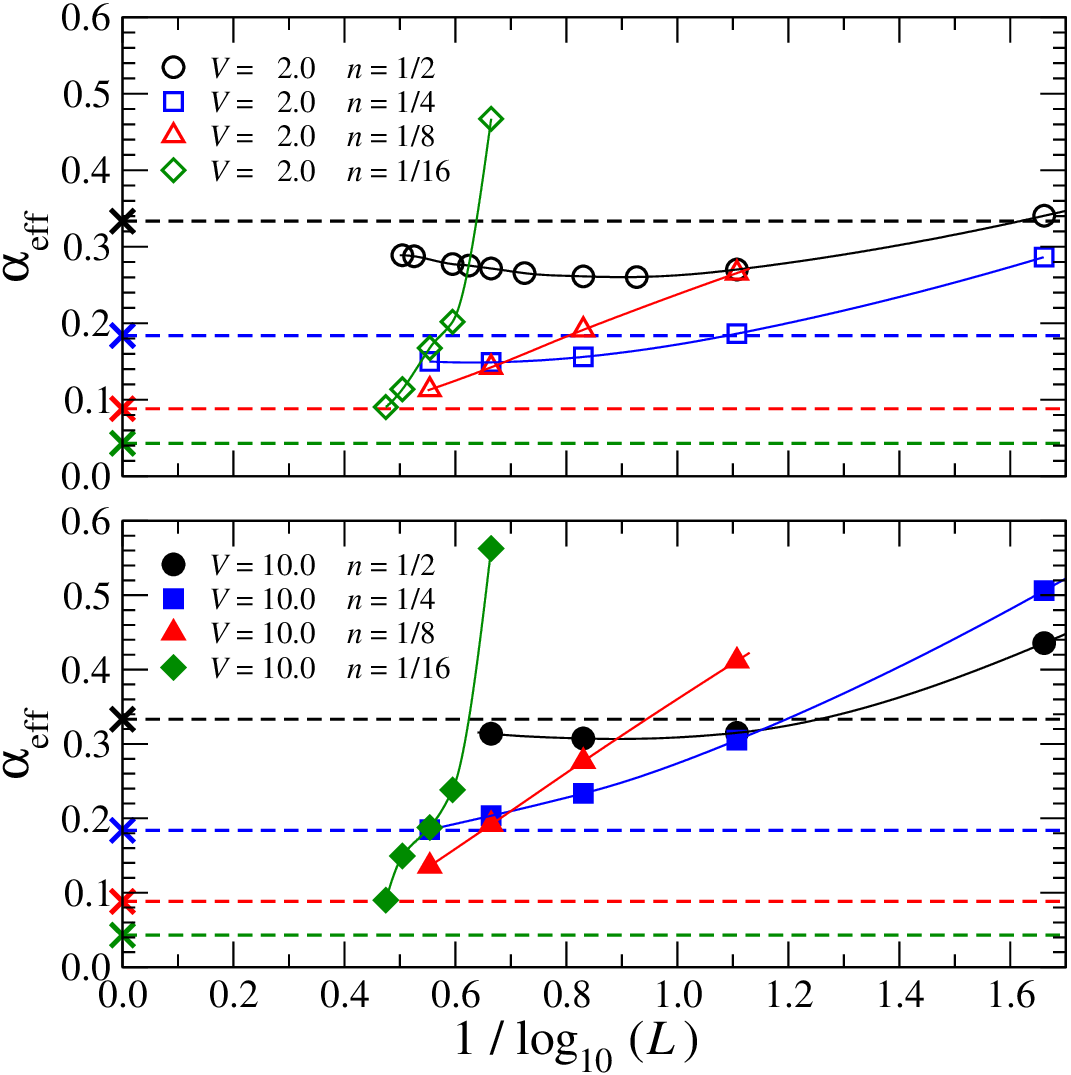}}
\caption{(Color online) The effective exponents versus the inverse
of $\log_{10}(L)$ for the data displayed in Fig.~\ref{pcgi}. The
horizontal dashed lines correspond to the asymptotic exponents
$\alpha_{\rm eff} (n)$ with $n$ as listed in the legend.}
\label{aegi}
\end{figure}

Studies of the persistent current at low fillings try to address the
question of differences between discrete lattice models and
continuous models of one-dimensional systems. Such questions have
been addressed in a recent comprehensive numerical
study~\cite{Nemeth, Krcmar}. Discrete models mimic continuous
electron gas models if the electron concentration $n$ is such that
the cosine dispersion of the discrete system can reasonably well
approximate the parabolic dispersion of the electron gas. Moreover,
lattice models  violate Galilean invariance~\cite{Schick}. In a
Galilean invariant model without impurity ($V=0$) the persistent
current should not depend on the interaction $U$.

First we check whether the interaction dependence of the current
really disappears at lower fillings: Figure~\ref{galinv} shows
persistent currents as a function of system size $L$ for  $U=0$,
$U=1$, and $U=2$ at various fillings $n=\frac{1}{2}$, $\frac{1}{4}$,
and $\frac{1}{8}$. The currents at half-filling (full symbols)
strongly depend on the electron-electron interaction, while the
electron gas is better approximated at quarter-filling (shaded
symbols). A negligibly small dependence on $U$ is found at
$n=\frac{1}{8}$ (open symbols) where Galilean invariance is almost
satisfied.

Figure~\ref{pcgi} shows the decay of the persistent current as a
function of the ring size $L$ for electron concentrations 
$n=\frac{1}{2}$, $\frac{1}{4}$, $\frac{1}{8}$, and $\frac{1}{16}$ on
a logarithmic scale for $V=2$ (open symbols) and $V=10$ (full symbols).
The asymptotic values $\alpha_{\rm B}(n)$ expected from the
Bosonization analysis~\cite{Haldane, Qin, And} are indicated by the
dashed straight lines corresponding to exponents $\alpha_{\rm
B}(n=\frac{1}{2})=\frac{1}{3}$, $\alpha_{\rm
B}(\frac{1}{4})=0.18383$, $\alpha_{\rm B}(\frac{1}{8})=0.08825$,
$\alpha_{\rm B}(\frac{1}{16})=0.042940$. On the scale of the figure
the data seemingly reach the asymptotics. A more detailed picture of
how the effective power-law exponents $\alpha_{\rm eff}$ approach
the expected asymptotic values $\alpha_{\rm B}$ is shown in
Fig.~\ref{aegi}. The upper and lower panels show the effective
exponent calculated for $V=2$ and $V=10$, respectively, at various
fillings $n$. It is obvious that the $\alpha_{\rm eff}(n)$ do
not converge to $\alpha_{\rm B}(n)$,
and substantially larger system sizes $L$ would need to considered
in order to see convergence. From our data we, therefore, cannot
support the suggestion~\cite{Nemeth, Krcmar} that the asymptotic
regime can be reached at a small number of electrons $nL$ in
continuous models.

\section{Summary}

We studied the asymptotic power law decay of persistent currents
using three different numerical methods (fRG, DMRG, ED). To this
end we improved the fRG method so that systems with up to
$3\cdot10^4$ sites could be treated. We used DMRG and ED in order to
benchmark the fRG results and confirmed a sufficiently good
agreement between fRG, DMRG, and ED calculations. We confirmed
that the fRG is a suitable method for extracting the asymptotic
Luttinger power law exponent $\alpha_{\rm B}$.

We found that $\alpha_{\rm B}$ which describes the decay of the
persistent current is about 5\% overestimated by the fRG results
compared to expectations from Bosonization and additional
approximation. This deviation we attribute to the approximation
of the fRG method, and the effective exponent is known to be
correct to the leading order in the electron-electron interaction
from the calculations of the same exponent for quantum wires.
The accurate DMRG analysis cannot, unfortunately, be extended
to sufficiently large systems. However, there are indications that
the expected asymptotic values could be reached, if larger systems
could be calculated. It is found that the effective exponent for a
fixed impurity strength $V$ shows a typical minimum for lengths
$8 \lesssim L \lesssim 100$ at half-filling.

Generally, it is known that the stronger the impurity, the faster
the asymptotic regime is reached. However, we identified three
regions for $U=1$: in the first region, $0<V\lesssim4$, we observed
the standard behavior. (The case $V=4$ corresponds to
$T=\frac{1}{5}$.) In the the second region $4\lesssim V\lesssim 10^3$
this behavior is reversed. In the third region $V>10^3$ the
dependence of the effective exponent $\alpha_{\rm eff}$ on the
impurity strength $V$ saturates.

Since discrete lattice models are not Galilean invariant, we
decreased the electron filling down to $n=\frac{1}{16}$ such that
the discrete Hamiltonian mimics a continuous electron gas with a
quadratic dispersion law. From our data we see that even
at low fillings, i.e. for quasi-continuous models, we need large
systems in order to reach the expected power law exponents.

\section*{Acknowledgments}
We thank S. Andergassen, V.~Meden, and U.~Schollw\"ock for stimulating
discussions. This work is partially supported by the Slovak Agency
for Science and Research grant APVV-51-003505, APVV-VVCE- 0058-07, QUTE,
and VEGA grant No. 1/0633/09 (A.G. and R.K.). A.G. also
acknowledges support from the Alexander von Humboldt foundation.
A.G. and R. K. thank PTB for hospitality and support.

\appendix
\section{Solution of the fRG equations for large $k$}
\label{initial cond}

Here we show how the initial conditions given in
Eqs.~(\ref{incondeff}) are determined. We start from the equations
(\ref{floweqs}) for the coefficients $a_{k}$ for large
$k$
\begin{equation}
a_{k,jj}^\prime=\frac{2U}{\pi }\frac{\sin(k~0^+)}{k},\quad a_{k,j(j\pm1)}^\prime=0,
\end{equation}
noting that $U_k=U$ for large $k$ ($0^+$ denotes a positive
infinitesimal increment). This equation is easily solved using the initial
conditions at $k=\infty$ given in Eq.~(\ref{initialcond}) with the
result
\begin{eqnarray}
a_{k,jj}&=&-U\left(1-\frac{2}{\pi}{\rm Si}(k~0^+)\right)\delta_{jj}-V\delta_{j1}+\mu\delta_{jj},\nonumber\\
\quad a_{k,j(j\pm1)}&=&0.
\end{eqnarray}
Here, ${\rm Si}(z)$ is the sine integral function. The impurity with
strength $V$ is placed at site $j=1$. This result can now be used in
order to solve the equation for $a_{k,0}$ for large k,
\begin{equation}\label{eqa0}
a_{k,0}^\prime=\frac{L\mu-V}{\pi}\frac{\sin(k~0^+)}{k}-\frac{LU}{\pi}\frac{\sin(k~0^+)}{k}\left(1-\frac{2}{\pi}{\rm Si}(k~0^+)\right)
\end{equation}
where the expansion  $\ln(1+x)\approx x$ for $x<1$ has been used.
Integration of Eq.~(\ref{eqa0})  yields
\begin{equation}
a_{k,0}=\frac{V}{2}+\frac{L}{2}\left(\frac{U}{2}-\mu\right).
\end{equation}
where the limit $0^+ \rightarrow 0$ has been performed.

\section{Numerical solution of the flow equations for the self energies
and the ground state energy}\label{numerics}

The solution of the flow equations for the self-energies and ground
state energy~(\ref{floweqs}) requires the calculation of the inverse
and determinant of (potentially) large cyclic tridiagonal matrices
at each step of the integration of a set of ordinary differential
equations. Of course, this could be achieved with standard library
routines. However, due to the special form of the
Eqs.~(\ref{floweqs}) we only need to determine the cyclic
tridiagonal part of these matrices. Therefore, in the following we
will develop methods adapted to our special needs. As a consequence
we achieve a significant speed-up of the numerical calculation as
well as considerable reduction of the computer memory requirements.
Such improvements enable us to treat large system sizes.

\subsection{Inversion of cyclic tridiagonal matrices}

For matrices $\mathbf{M}$ which can be represented as the sum of a
tridiagonal matrix $\mathbf{T}$ and a direct product of two vectors
$\mathbf{u}$ and $\mathbf{v}$,
\begin{equation}
\label{sher}
\mathbf{M} = \mathbf{T} + \mathbf{u}\otimes \mathbf{v},
\end{equation}
the inverse can be obtained using the Sherman-Morrison formula
\cite{rec}
\begin{equation}\label{shermor}
\mathbf{M}^{-1}=\mathbf{T}^{-1} - \frac{\mathbf{z}\otimes \mathbf{w}}{1 + \mathbf{v}\cdot \mathbf{z}}
\end{equation}
with
\begin{equation}\label{zw}
 \mathbf{z}=\mathbf{T}^{-1} \mathbf{u},  \quad\quad\quad
\mathbf{w}=(\mathbf{T}^{-1})^T\mathbf{v}.
\end{equation}

A general cyclic tridiagonal matrix
\begin{equation}
 \mathbf{M}=\left(
\begin{array}{ccccc}
a_1    & b_1    & 0      &  \cdots  & c_1 \\
c_2    & a_2    & b_2    &          & 0      \\
0      & \ddots &\ddots  &  \ddots  & \vdots \\
\vdots &        &c_{n-1} & a_{n- 1} & b_{n-1}\\
b_n & \cdots &0       & c_{n}  & a_n
\end{array}
\right)
\end{equation}
can be written in the form (\ref{sher}) with the tridiagonal part
$\mathbf{T}$ given by
\begin{equation}
\mathbf{T}=\left(
\begin{array}{ccccc}
\alpha_1    & b_1    & 0      &  \cdots  & 0  \\
c_2    & \alpha_2    & b_2    &          & 0      \\
0      & \ddots &\ddots  &  \ddots  & \vdots \\
\vdots &        &c_{n-1} & \alpha_{n- 1} & b_{n-1}\\
0 & \cdots &0       & c_{n}  & \alpha_n
\end{array}
\right)
\end{equation}
with $\alpha_1=2a_1$, $\alpha_n=a_n+c_1b_n/a_1$ and $\alpha_i=a_i$
for $1<i<n$. The vectors $\mathbf{u}$ and $\mathbf{v}$ are defined
by
\begin{equation}
 \mathbf{u}=\left(-a_1, 0, \cdots, 0,  b_n \right), \quad \mathbf{v}=\left(1, 0, \cdots, 0, -c_1/a_1
\right).
\end{equation}

For the inversion of the tridiagonal part $\mathbf{T}$, we use a
method inspired by Andergassen {\it et al.}~\cite{And}, which is
based on Ref.~\cite{meu}. However, in our case the tridiagonal
matrix is not complex symmetric, therefore, we have to use an
$\mathbf{LU}$ decomposition of $\mathbf{T}$ instead of the
$\mathbf{LDU}$ decomposition employed in Ref.~\cite{And}. Details
are described in the following subsection.

\subsection{Implementation}
According to Eq.~(\ref{shermor}) the inversion of a general cyclic
tridiagonal matrix requires two essential steps:  {\it (i)} the
inversion of a tridiagonal matrix and  {\it (ii)} the determination
of the vectors $\mathbf{z}$ and $\mathbf{w}$.

{\it (i) } A tridiagonal matrix can be represented as a product of
two matrices
$\mathbf{T}=\mathbf{LU}=\tilde{\mathbf{U}}\tilde{\mathbf{L}}$, where
$\mathbf{U}$, $\tilde{\mathbf{U}}$ are upper diagonal and
$\mathbf{L}$, $\tilde{\mathbf{L}}$ are lower diagonal matrices. From
the $\mathbf{L}\mathbf{U}$ decomposition
\begin{eqnarray}
&&\mathbf{T} = \mathbf{L} \mathbf{U}=\\
&&\left(
\begin{array}{ccccc}
d_1 &  &   & & \\
c_2 & d_2 &  &    \\
  &\ddots &\ddots& & \\
 &  &  c_{n-1} & d_{n- 1} & \\
  & & & c_n& d_n
\end{array}\right)
\left(
\begin{array}{ccccc}
1 & u_1 &   & & \\
 & 1 & u_2 &    \\
  & &\ddots&\ddots& \\
 &  &   & 1 & u_{n- 1}\\
  & & & & 1
\end{array}
\right)\nonumber
\end{eqnarray}
with
\begin{equation}
 d_1=\alpha_1,\quad d_i = \alpha_i-u_{i-1}c_i, \quad u_i=b_i/d_i
\end{equation}
one easily finds  for $ 1<i<j<n$ the  relations
\begin{equation}\label{UL}
 (\mathbf{T}^{-1})_{n,n}=1/d_n,\quad (\mathbf{T}^{-1})_{i,j} = -u_{i}(\mathbf{T}^{-1})_{i+1,j}.
\end{equation}
Similarly, from the $\tilde{\mathbf{U}}\tilde{\mathbf{L}}$
decomposition of $\mathbf{T}$,
\begin{eqnarray}
&&\mathbf{T}\!=\!\tilde{\mathbf{U}}\tilde{\mathbf{L}}=\\
&&\left(\begin{array}{ccccc}
1 & \tilde{u}_1 &   & & \\
 & 1 &\tilde{u}_2  &    \\
  & &\ddots&\ddots & \\
 &  &   & 1 &\tilde{u}_{n-1} \\
  & & & & 1
\end{array}
\right)\!\!\left(
\begin{array}{ccccc}
\tilde{d}_1 &  &   & & \\
c_2 & \tilde{d}_2 &  &    \\
  &\ddots &\ddots&& \\
 &  & c_{n-1}  & \tilde{d}_{n-1} & \\
  & & &c_n & \tilde{d}_n
\end{array}
\right)\nonumber
\end{eqnarray}
with
\begin{equation}
 \tilde{d}_n=\alpha_n \quad \tilde{d}_i = \alpha_i - \tilde{u}_i c_{i+1} \quad \tilde{u}_i = b_i/\tilde{d}_{i+1},
\end{equation}
one obtains for $1<i<j<n$ the relations
\begin{equation}\label{LU}
\label{1r}
 (\mathbf{T}^{-1})_{1,1}=1/\tilde{d}_1,\quad  (\mathbf{T}^{-1})_{i,j}=-\tilde{u}_{j-1}(\mathbf{T}^{-1})_{i,j-1}.
\end{equation}

Combining Eqs.~(\ref{UL}) and (\ref{LU}) yields a recursion relation
for the diagonal elements of $\mathbf{T}^{-1}$
\begin{equation}
 \label{diag}
(\mathbf{T}^{-1})_{i+1,i+1}=\frac{\tilde{u}_i}{u_i}(\mathbf{T}^{-1})_{i,i}
\end{equation}
with the initial condition given in Eq.~(\ref{LU}). From the
diagonal elements one then generates the upper diagonal as well as
the upper corner element using the relations given above.
Alternatively, the $\tilde{\mathbf{U}}\tilde{\mathbf{L}}$
decomposition yields the relation
\begin{equation}
 (\mathbf{T}^{-1})_{i,j} = -\frac{c_{i}}{\tilde{d}_{i}}(\mathbf{T}^{-1})_{i-1,j}~~~~{\rm for}~ 1<j<i<n,
\end{equation}
which is used to generate the lower diagonal and corner element,
respectively.

{\it (ii)} The vector $\mathbf{z}$ is obtained from Eq.~(\ref{zw}),
$\mathbf{Tz}=\mathbf{LUz}=\mathbf{u}$, as follows: first we solve
recursively $\mathbf{L}\mathbf{z}^\prime=\mathbf{u}$ for
$\mathbf{z}^\prime$
\begin{equation}
 z_1^\prime = u_1/d_1, \quad z^\prime_i = (u_i - c_iz^\prime_{i-1})d_i~~~~{\rm for}~ 1<i\leq n,
\end{equation}
and then obtain $\mathbf{z}$ recursively from $\mathbf{U}\mathbf{z}
= \mathbf{z}^\prime$
\begin{equation}
  z_n = z_n^\prime, \quad z_i = z_i^\prime - u_iz_{i+1}~~~~{\rm for}~ 1\leq i <n.
\end{equation}
In a similar way one determines $\mathbf{w}$ from
$\mathbf{T}^T\mathbf{w}=\mathbf{v}$.

\subsection{Determinant of a cyclic tridiagonal matrix}

The determinant of a cyclic tridiagonal matrix is calculated using a
formula given in Ref.~\cite{moli}
\begin{eqnarray}
& &\det\left(
\begin{array}{ccccc}
a_1    & b_1    & 0      &  \cdots  & c_1 \\
c_2    & a_2    & b_2    &          & 0      \\
0      & \ddots &\ddots  &  \ddots  & \vdots \\
\vdots &        &c_{n-1} & a_{n- 1} & b_{n-1}\\
b_n & \cdots &0       & c_{n}  & a_n
\end{array}
\right) = \\
\nonumber\\
& &\quad(-1)^{n + 1}\left(\prod_{i=1}^n b_i + \prod_{i=1}^n c_i\right)+
   {\rm Tr}\prod_{i=1}^n\left(  \begin{array}{cc}
    a_i & -b_{i - 1}c_i\\
    1   &  0           \end{array}\right)\nonumber
\end{eqnarray}
with $b_0=b_n$. For large systems, the products in this formula can
easily underflow or overflow. This must be carefully controlled
appropriately.

\section*{References}

\bibliography{paper2}

\begin{thebibliography}{10}

\bibitem{Deaver}
B.S.~Deaver Jr. and W.~M. Fairbank.
\newblock {\em Phys. Rev. Lett.}, 7:43, 1961.

\bibitem{Byers}
N.~Byers and C.~N. Yang.
\newblock {\em Phys. Rev. Lett.}, 7:46, 1961.

\bibitem{Buttiker}
M.~B\"{u}ttiker, Y.~Imry, and R.~Landauer.
\newblock {\em Phys. Lett.}, 96A:365, 1983.

\bibitem{Levy}
L.P. L\'{e}vy, G.~Dolan, J.~Dunsmuir, and H.~Bouchiat.
\newblock {\em Phys. Rev. Lett.}, 64:2074, 1990.

\bibitem{Chandrasekhar}
V.~Chandrasekhar, R.~A. Webb, M.~J. Brady, M.~B. Ketchen, W.~J. Gallagher~W J,
  and A.~Kleinsasser.
\newblock {\em Phys. Rev. Lett.}, 67:3578, 1991.

\bibitem{Mailly}
D.~Mailly, C.~Chapelier, and A.~Benoit.
\newblock {\em Phys. Rev. Lett.}, 70:2020, 1993.

\bibitem{Reulet}
B.~Reulet, M.~Ramin, H.~Bouchiat, and D.~Mailly D.
\newblock {\em Phys. Rev. Lett.}, 75:124, 1995.

\bibitem{Jariwala}
E.~M.~Q. Jariwala, P.~Mohanty, M.~B. Ketchen, and R.~A. Webb.
\newblock {\em Phys. Rev. Lett.}, 86:1594, 2001.

\bibitem{Rabaut}
W.~Rabaout, L.~Saminadayar, D.~Mailly, K.~Hasselbach, A.~Beno$\hat{\rm i}$t,
  and B.~Etienne.
\newblock {\em Phys. Rev. Lett.}, 86:3124, 2001.

\bibitem{Kleemans}
N.~A. J. M.~Kleemans et~al.
\newblock {\em Phys. Rev. Lett.}, 99:146808, 2007.

\bibitem{Bouchiat}
H.~Bouchiat.
\newblock {\em Physics}, 1:7, 2008.

\bibitem{Landauer}
R.~Landauer and M.~B\"{u}ttiker.
\newblock {\em Phys. Rev. Lett.}, 54:2049, 1985.

\bibitem{Cheung}
H.~F. Cheung, Y.~Gefen, E.~K. Riedel, and W.~H. Shih.
\newblock {\em Phys. Rev. B}, 37:6050, 1988.

\bibitem{Abraham}
M.~Abraham and R.~Berkovits.
\newblock {\em Phys. Rev. Lett.}, 70:1509, 1993.

\bibitem{Bouzerar}
G.~Bouzerar, D.~Poiblanc, and G.~Montambaux.
\newblock {\em Phys. Rev. B}, 49:8258, 1994.

\bibitem{Gogolin}
A.~O. Gogoglin and Prokof'ev.
\newblock {\em Phys. Rev. B}, 50:4921, 1994.

\bibitem{Kato}
H.~Kato and D.~Yoshioka.
\newblock {\em Phys. Rev. B}, 50:4943, 1994.

\bibitem{Cohen}
A.~Cohen, K.~Richter, and R.~Berkovits.
\newblock {\em Phys. Rev. B}, 57:6223, 1998.

\bibitem{Henkel}
M.~Henkel and D.~Karevski.
\newblock {\em Eur. Phys. J. B}, 5:787, 1998.

\bibitem{Jaimungal}
S.~Jaimungal, M.~H.~S. Amin, and G.~Rose.
\newblock {\em Int. J. Mod. Phys. B}, 13:3171, 1999.

\bibitem{Meden}
V.~Meden and U.~Schollw\"{o}ck.
\newblock {\em Phys. Rev. B}, 67:035106, 2003.

\bibitem{Enss}
T.~Enss, V.~Meden, S.~Andergassen, X.~Barnab\'{e}-Th\'{e}riault, W.~Metzner,
  and K.~Sch\"{o}nhammer.
\newblock {\em Phys. Rev. B}, 71:155401, 2005.

\bibitem{Friedrich}
S.~Friederich and V.~Meden.
\newblock {\em Phys. Rev. B}, 77:195122, 2008.

\bibitem{Soroker}
H.~Bary-Soroker, O.~Entin-Wohlmann, and Y.~Imry.
\newblock {\em Phys. Rev. Lett.}, 101:057001, 2008.

\bibitem{Heinrichs}
J.~Heinrichs.
\newblock {\em J. Phys.: Condens. Matter}, 20:345232, 2008.

\bibitem{Altshuler}
B.~L. Altshuler, Y.~Gefen, and Y.~Imry Y.
\newblock {\em Phys. Rev. Lett.}, 66:88, 1991.

\bibitem{Ambegaokar}
V.~Ambegaokar and U.~Eckern.
\newblock {\em Europhys. Lett.}, 13:733, 1990.

\bibitem{Cesare}
E.~Gambetti-C\'{e}sare, D.~Weinmann, R.~A. Jalabert, and Ph. Brune.
\newblock {\em Eurphys. Lett.}, 60:120, 2005.

\bibitem{Gambetti}
E.~Gambetti.
\newblock {\em Phys. Rev. B}, 72:165338, 2005.

\bibitem{Krcmar}
R.~Kr\v{c}m\'{a}r, A.~Gendiar, M.~Mo\v{s}ko, R.~N\'{e}meth, P.~Vagner, and
  L.~Mitas.
\newblock {\em Physica E}, 40:1507, 2008.

\bibitem{Nemeth}
R.~N\'{e}meth, M.~Mo\v{s}ko, R.~Kr\v{c}m\'{a}r, A.~Gendiar, M.~Indlekofer, and
  L.~Mitas.
\newblock {\em arXiv:0902.2225}.

\bibitem{Luther}
A.~Luther and I.~Peschel.
\newblock {\em Phys. Rev. B}, 9:2911, 1974.

\bibitem{Apel}
W.~Apel and T.~M. Rice.
\newblock {\em Phys. Rev. B}, 26:7063, 1982.

\bibitem{Kane1}
C.~L. Kane and M.~P.~A. Fisher.
\newblock {\em Phys. Rev. Lett.}, 68:1220, 1992.

\bibitem{Kane2}
C.~L. Kane and M.~P.~A. Fisher.
\newblock {\em Phys. Rev. B}, 46:15233, 1992.

\bibitem{Haldane}
F.~D.~M. Haldane.
\newblock {\em Phys. Rev. Lett.}, 45:1358, 1980.

\bibitem{Matveev}
K.~A. Matveev, D.~Yue, and L.~I. Glazmann.
\newblock {\em Phys. Rev. Lett.}, 71:3351, 1993.

\bibitem{Glazmann}
D.~Yue, L.~I. Glazmann, and K.~A. Matveev.
\newblock {\em Phys. Rev. B}, 49:1966, 1994.

\bibitem{Polchinski}
J.~Polchinski.
\newblock {\em Nucl. Phys. B}, 231:269, 1984.

\bibitem{Wetterich}
C.~Wetterich.
\newblock {\em Phys. Lett. B}, 301:90, 1993.

\bibitem{Morris}
T.~R. Morris.
\newblock {\em Int. J. Mod. Phys. A}, 9:2411, 1994.

\bibitem{Salmhofer1}
M.~Salmhofer.
\newblock {\em Commun. Math. Phys.}, 194:249, 1998.

\bibitem{Salmhofer2}
M.~Salmhofer and C.~Honerkamp.
\newblock {\em Prog. Theor. Phys.}, 105:1, 2001.

\bibitem{White1}
S.~R. White.
\newblock {\em Phys. Rev. Lett.}, 69:2863, 1992.

\bibitem{White2}
S.~R. White.
\newblock {\em Phys. Rev. B}, 48:10345, 1993.

\bibitem{Peschel}
I.~Peschel, X.~Wang, M.~Kaulke, and K.~Hallberg (Eds.).
\newblock {\em Lecture Notes in Physics {\bf 528} Density-Matrix
  Renormalization, A New Numerical Method in Physics}.
\newblock Springer, Berlin, 1999.

\bibitem{Scholl}
U.~Schollw\"{o}ck.
\newblock {\em Rev. Mod. Phys.}, 77:259, 2005.

\bibitem{Peierls}
R.~E. Peierls.
\newblock {\em Z. Phys. B: Condens. Matter}, 80:763, 1933.

\bibitem{SCH98}
V.~Meden, P.~Schmitteckert, and N.~Shannon.
\newblock {\em Phys. Rev. B}, 57:8878, 1998.

\bibitem{And}
S.~Andergassen, T.~Enss, V.~Meden, W.~Metzner, U.Schollw\"ock, and
  K.~Sch\"onhammer.
\newblock {\em Phys. Rev. B}, 70:075102, 2004.

\bibitem{MED02a}
V.~Meden, W.~Metzner, U.~Schollw{\"o}ck, and K.~Sch{\"o}nhammer.
\newblock {\em Phys. Rev. B}, 65:045318, 2002.

\bibitem{MED02b}
V.~Meden, W.~Metzner, U.~Schollw{\"o}ck, and K.~Sch{\"o}nhammer.
\newblock {\em J. Low Temp. Phys.}, 126:1147, 2002.

\bibitem{AND04}
S.~Andergassen, T.~Enss, V.~Meden, W.~Metzner, U.~Schollw{\"o}ck, and
  K.~Sch{\"o}nhammer.
\newblock {\em Phys. Rev. B}, 70:075102, 2004.

\bibitem{WEY08}
M.~Weyrauch and D.~Sibold.
\newblock {\em Phys. Rev. B}, 77:125309, 2008.

\bibitem{BER00}
J.~Berges, N.~Tetradis, and C.~Wetterich.
\newblock {\em Phys. Rep.}, 363:223--386, 2002.

\bibitem{Schick}
M.~Schick.
\newblock {\em Phys. Rev.}, 166:144, 1967.

\bibitem{Qin}
S.~Qin, M.~Fabrizio, L.~Yu, M.~Oshikawa, and I.~Affleck.
\newblock {\em Phys. Rev. B}, 56:9766, 1997.

\bibitem{rec}
S.A. Teukolsky, W.~T. Vetterling, and B.P.Flannery.
\newblock {\em Numerical Recipes in Fortran 77}.
\newblock Cambrige University Press, Cambridge, MA, 1986.

\bibitem{meu}
G.~Meurant.
\newblock {\em SIAM J. Matrix Anal. Appl.}, 13:707, 1992.

\bibitem{moli}
L.~G. Molinari.
\newblock {\em arXiv:0712.0681v3}.

\end{thebibliography}
\bibliographystyle{unsrt}

\end{document}